\documentclass[psfig,showpacs,twocolumn,prl]{revtex4}
\usepackage{graphicx}

\begin{document}

\title{Quantum phase transitions in a spin-$1/2$ alternating Heisenberg
antiferromagnetic chain under a staggered transverse magnetic field}
\author{Guang-Qiang Zhong, Shou-Shu Gong, Qing-Rong Zheng, Gang Su$^{\ast}$}
\affiliation{College of Physical Sciences, Graduate University of Chinese Academy of
Sciences, P. O. Box 4588, Beijing 100049, China}

\begin{abstract}
The magnetic behaviors of a spin-$1/2$ alternating Heisenberg
antiferromagnetic chain in a staggered transverse magnetic field is studied
by means of the density-matrix renormalization group method and
Jordan-Wigner transformation. Quantum phase transitions of different types
are observed in the $S$=$1$ N\'{e}el and XY-like gapless phases, which
result from the competitions between the staggered transverse field and
magnetic orders induced by anisotropy and alternating interactions. The
results are compared with the mean-field and some exactly resolved results.
\end{abstract}

\pacs{75.10.Jm, 75.40.Cx}
\maketitle

One-dimensional quantum magnets have been at the center of theoretical and
experimental attention due to the exotic magnetic properties in condensed
matter physics. In particular, the magnetic behaviors of low dimensional
quantum magnets in an external magnetic field exhibit many novel
characteristics \cite{group}. The gapped Haldane chain \cite{Haldane}
compound Ni(C$_{2}$H$_{8}$N$_{2}$)$_{2}$NO$_{2}$(ClO$_{4}$) in a uniform
magnetic field has been widely studied both experimentally \cite%
{uniform1,uniform2} and theoretically \cite{uniform3}, which shows a
commensurate-incommensurate transition as the gap is closed by the field
\cite{CG}. A $S$=$1$ Haldane chain compound R$_{2}$BaNiO$_{5}$ (R=magnetic
rare earth) \cite{ZRM,MZ} with an effective staggered magnetic field has
been explored theoretically \cite{MZ2,EM} and numerically \cite{LD},
revealing that the magnetic behavior in a staggered field is totally
different from that in a uniform field. Recently, the magnetic properties of
a $S$=$1/2$ antiferromagnetic (AF)-ferromagnetic (FM) spin chain are
extensively studied \cite{Sakai,MA,AM}. This spin chain with nearly the same
AF and FM interaction strength has been realized in experiment by the
compound (CH$_{3}$)$_{2}$NH$_{2}$CuCl$_{3}$ \cite{ST}. Hida \cite{Hida}
pointed out that this chain can map onto the $S$=$1$ Haldane chain when the
FM couplings dominate. Yamanaka \textit{et al.} \cite{YHK} suggested a phase
diagram for the system with an AF anisotropy. The system has the Haldane, $S$%
=$1$ N\'{e}el, and XY-like gapless phases for different anisotropies and
alternations. In the $S$=$1$ N\'{e}el and XY-like phases, the gap vanishes
and some magnetic orders emerge. In this paper, we shall concentrate
primarily on the magnetic properties of the system under a staggered
transverse field in various phases. It is expected that the competitions
between different factors would yield rich results.

Let us consider a spin-1/2 alternating Heisenberg chain with anisotropy in a
transverse staggered magnetic field, as depicted in the inset of Fig. \ref%
{phase}. The Hamiltonian reads
\begin{eqnarray}
H &=&\sum_{j=1}^{N}(S_{2j-1}^{x}S_{2j}^{x}+S_{2j-1}^{y}S_{2j}^{y}+\lambda
S_{2j-1}^{z}S_{2j}^{z})  \nonumber \\
&+&\beta \sum_{j=1}^{N}\vec{S}_{2j}\cdot \vec{S}_{2j+1}-h_{s}%
\sum_{j=1}^{2N}(-1)^{j}S_{j}^{x},  \label{Hamiltonian}
\end{eqnarray}%
where $\lambda $ stands for the AF anisotropy, $\beta <0$ ($>0$) is the FM
(AF) coupling, $N$ is the number of unit cells, and the last term of Eq. (%
\ref{Hamiltonian}) introduces the staggered transverse magnetic field. We
take the AF coupling as the energy scale and $g\mu _{B}$=$1$. The transverse
staggered magnetization and susceptibility are defined, respectively, as
\begin{eqnarray}
m_{stag} &=&\frac{1}{2N}\sum_{j=1}^{2N}(-1)^{j}\langle S_{j}^{x}\rangle , \\
\chi _{stag} &=&\frac{\partial {m_{stag}}}{\partial {h_{s}}}.
\end{eqnarray}

The density-matrix renormalization group (DMRG) method \cite{DMRG1,DMRG2} is
invoked to study this system. In our calculations, the length of the chain
is taken as $60$, and the number of the optimal states is kept as $60$. We
adopt open boundary conditions. The truncation error is less than $10^{-3}$
in all calculations. The system has an U(1) symmetry in absence of magnetic
field, however, the staggered transverse field breaks this symmetry. Thus,
there is no good quantum number that can be used to reduce the Hilbert space
dimensions in our calculations.

\begin{figure}[tbp]
\includegraphics[width = 0.6\linewidth,clip]{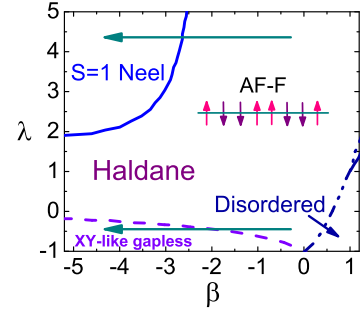}
\caption{(color online) Phase diagram of the spin-$1/2$ AF-FM alternating
quantum spin chain with Hamiltonian (\protect\ref{Hamiltonian}) \protect\cite%
{YHK}. The staggered magnetic behaviors along the arrows will be studied in
the context.}
\label{phase}
\end{figure}

In order to investigate the effects of the anisotropy and FM interaction on $%
m_{stag}$, we first calculate exactly $m_{stag}$ of two $S$=$1/2$ spins
coupled by XXZ AF and isotropic FM interactions, respectively. For the spins
coupled by XXZ AF coupling, $m_{stag}$ behaves as
\begin{equation}
m_{stag}^{AF}=\frac{2h_{s}}{\sqrt{16h_{s}^{2}+(1+\lambda)^{2}}}.
\end{equation}
It can be seen that with increasing $\lambda$, $m_{stag}$ is supressed by
the anisotropy that destructs the transverse magnetic ordering. For the
spins coupled by the FM interaction, $m_{stag}$ behaves as
\begin{equation}
m_{stag}^{FM}=\frac{h_{s}}{\sqrt{4h_{s}^{2}+\beta^{2}}}.
\end{equation}
Clearly, the FM interaction makes the spins align in the same direction,
which competes with the staggered magnetic field. Thus, $m_{stag}$ is
supressed with increasing $\beta$. It seems that the anisotropy and FM
interaction have similar effects on $m_{stag}$. However, when the FM
coupling and the staggered magnetic field are considered simultaneously, $%
m_{stag}$ shows more complex behaviors. In the following, the transverse
staggered magnetic properties will be studied by means of the DMRG method
for the parameters indicated by the arrows of Fig. \ref{phase}.

\begin{figure}[tbp]
\includegraphics[width = 1.0\linewidth,clip]{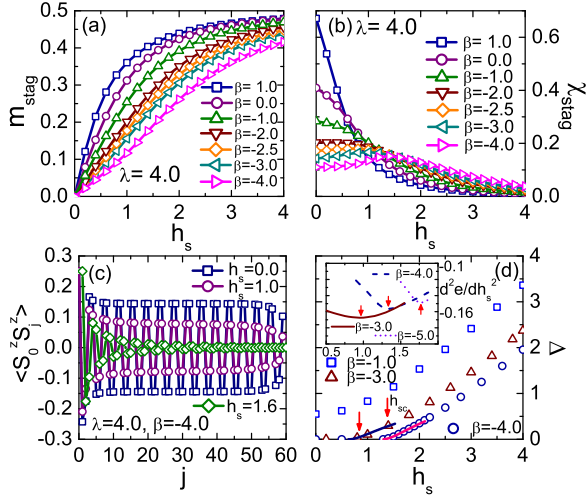}
\caption{(color online) Field dependence of $m_{stag}$, $\protect\chi_{stag}$%
, $\langle S^{z}_{0}S^{z}_{j}\rangle$, and $\Delta$ for $\protect\lambda$=$%
4.0$ in (a), (b), (c), and (d), respectively. The lines in (d) denote the
fitted results with Eq. (\protect\ref{fit1}). The inset of (d) shows the
second derivative of the ground-state energy $\partial^{2}{e}/\partial{h_{s}}%
^{2}$ as a function of $h_{s}$.}
\label{H-N}
\end{figure}

Figure \ref{H-N} shows the evolution of the magnetic properties from the
Haldane phase to the $S$=$1$ N\'{e}el phase with changing $\beta$ from $\beta
$=$1$ to $\beta$=$-4$ for $\lambda$=$4.0$. In Fig. \ref{H-N}(a), $m_{stag}$
increases with $h_{s}$ and approaches $0.5$ when $h_{s}$$\rightarrow$$\infty$%
. With increasing FM interactions, $m_{stag}$ declines for any field, which
agrees with the depressing effect of the FM coupling. In the Haldane phase, $%
\chi_{stag}$ continuously declines with $h_{s}$ in a simple way
\begin{equation}
\chi_{stag}\sim\frac{1}{(h_{s}^{2}+A)^{3/2}},
\end{equation}
where $A$ is a constant determined by the couplings. At $h_{s}$=$0$, the
zero-field susceptibility $\chi_{stag}(0)$ is finite, which is analogous to
the $S$=$1$ Haldane chain \cite{ZRM,MZ,MZ2,LD} due to the existence of a
gap. For the $S$=$1$ Haldane chain, $\chi_{stag}(0)$=$Z\upsilon/\Delta^{2}$ (%
$\Delta$ is the spin gap) \cite{MZ2}. In the present case, the spin gap \cite%
{Hida} and $\chi_{stag}(0)$ decrease simultaneously with increasing FM
coupling. Therefore, it is expected that the product of the renormalization
parameter $Z$ and the spin-wave velocity $\upsilon$ decrease more rapidly
than the gap. With further increasing the FM coupling, $\chi_{stag}$ becomes
flat at low fields when approaching the phase boundary. In the $S$=$1$ N\'{e}%
el phase, a broad peak emerges in $\chi_{stag}$ [Fig. \ref{H-N}(b)],
indicating the distinct magnetic properties below the transition field from
those in the Haldane phase. The peaks, which move to higher fields with
increasing $-\beta$, is attributed to the competition between the field and
magnetic interactions. In the absence of magnetic field, the longitudinal
spin-spin correlation function $\langle S^{z}_{0}S^{z}_{j}\rangle$ has a
long-range order (LRO) and behaves as $\langle S^{z}_{0}S^{z}_{2j-1}\rangle$$%
\simeq$$\langle S^{z}_{0}S^{z}_{2j}\rangle$ due to the FM coupling [Fig. \ref%
{H-N}(c)]. The N\'{e}el order in the $z$ axis prevents the magnetization in
the $x$ direction and is destructed by the transverse staggered field. As
shown in Fig. \ref{H-N}(c), with increasing $h_{s}$, $\langle
S^{z}_{0}S^{z}_{j}\rangle$ decays with a power law, and when $h_{s}$ exceeds
a critical magnetic field, $h_{s_{c}}$, it decays exponentially. Below $%
h_{s_{c}}$, the suppressed N\'{e}el order facilitates the magnetization in
the transverse axis, yielding the increase of $\chi_{stag}$. Above $h_{s_{c}}
$, the N\'{e}el order is fully broken and thus $\chi_{stag}$ declines
similar to that in the Haldane phase. In the Haldane phase, the gap enlarges
with increasing $h_{s}$. In the N\'{e}el phase, a gap is opened when $h_{s}$
exceeds $h_{s_{c}}$ and increases with the field, as shown in Fig. \ref{H-N}%
(d). In the transition field $h_{s_{c}}$, a quantum phase transition (QPT)
\cite{QPT} which is from the $S$=$1$ N\'{e}el phase to a gapped staggered
magnetic ordered phase happens. In the vicinity of $h_{s_{c}}$, the
ground-state energy $e$ and its derivatives with respect to the field are
studied to characterize this transition. It is found that both $\partial{e}%
/\partial{h_{s}}$ and $\partial^{2}{e}/\partial{h_{s}}^{2}$ are continuous
and nonsingular in the field $h_{s_{c}}$, but $\partial^{2}{e}/\partial{h_{s}%
}^{2}$ has a minimum at $h_{s_{c}}$, as shown in the inset of Fig. \ref{H-N}%
(d). Below $h_{s_{c}}$, the phase is gapless with a power law decaying $%
\langle S^{z}_{0}S^{z}_{j}\rangle$. Above $h_{s_{c}}$, a gap emerges and $%
\langle S^{z}_{0}S^{z}_{j}\rangle$ decays exponentially. This is analogous
to the transition from the XY-like phase to the Haldane phase in the absence
of the field \cite{KNO}, which is of the Kosterlitz-Thouless (KT) \cite{KT}
type. By considering the nonsingularity of the derivatives of $e$, we argue
that this QPT may be also of the KT type. The critical behavior of $\Delta$
near $h_{s_{c}}$ is fitted by the KT type with the lines in Fig. \ref{H-N}%
(d)
\begin{equation}
\Delta=De^{-C/\sqrt{h_{s}-h_{s_{c}}}}.  \label{fit1}
\end{equation}
For $\beta$=$-3(-4)$, $D$=$2.7(3.4)$, $C$=$2.0(2.0)$, $h_{s_{c}}$=$0.7$%
(1.27). The behavior of $\Delta$ near $h_{s_{c}}$ can be fitted well by Eq. (%
\ref{fit1}).

\begin{figure}[tbp]
\includegraphics[width = 1.0\linewidth,clip]{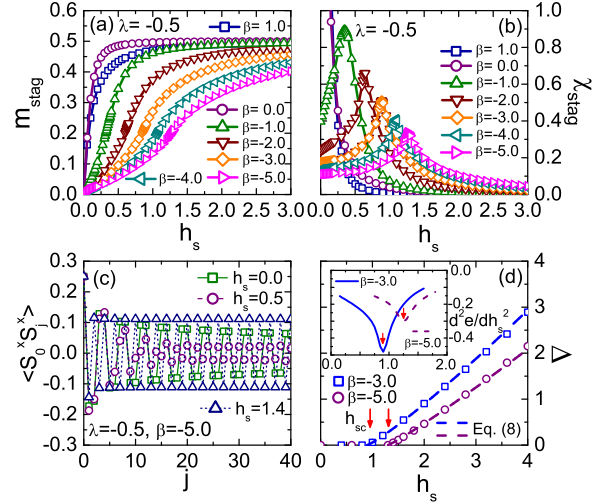}
\caption{(color online) Field dependence of $m_{stag}$, $\protect\chi_{stag}$%
, $\langle S^{x}_{0}S^{x}_{j}\rangle$, and $\Delta$ for $\protect\beta$=$-0.5
$ in (a), (b), (c), and (d), respectively. The inset of (d) shows $%
\partial^{2}{e}/\partial{h_{s}}^{2}$ as a function of $h_{s}$.}
\label{H-X}
\end{figure}

Figure \ref{H-X} shows the changes of magnetic properties from the Haldane
phase to the XY-like gapless phase with changing $\beta$ from $\beta$=$1$ to
$\beta$=$-5$ for $\lambda$=$-0.5$. In the XY-like gapless phase, $m_{stag}$
has an inflexion with increasing $h_{s}$, which is explicitly characterized
by the sharp peak of $\chi_{stag}$ and indicates a transition of magnetic
properties. Like the N\'{e}el phase, the peak moves to higher fields with
increasing $-\beta$ [Fig. \ref{H-X}(b)]. This transition is due to the
competition between the magnetic field and transverse magnetic ordering. In
the absence of magnetic field, $\langle S^{x}_{0}S^{x}_{j}\rangle$ has a
quasi-LRO, and due to the FM interaction, $\langle
S^{x}_{0}S^{x}_{2j-1}\rangle$$\simeq$$\langle S^{x}_{0}S^{x}_{2j}\rangle$,
i.e., $\langle S^{x}_{0}S^{x}_{j}\rangle$ has a translation symmetry with a
period of $4$ [Fig. \ref{H-X}(c)]. As the staggered field competes with the
FM interaction, this transverse quasi-LRO prevents the staggered
magnetization, and meanwhile, is destructed by increasing the field. With
increasing $h_{s}$, $\langle S^{x}_{0}S^{x}_{j}\rangle$ becomes disordered
in short range but builds an order with a $2$-period translation symmetry in
long range, as shown in Fig. \ref{H-X}(c) for $h_{s}$=$0.5$. When $h_{s}$
exceeds the transition field, the short-range disorder is replaced by the
staggered magnetic ordering, as in the case of $h_{s}$=$1.4$ in Fig. \ref%
{H-X}(c). These variations are also visible in spin static structure factor $%
S(q)$, which are not presented here for concise. In the absence of magnetic
field, $S(q)$ has two peaks at $q$=$\pi/2$ and $3\pi/2$. With increasing $%
h_{s}$, the old peaks decline rapidly and a new peak emerges at $q$=$\pi$,
indicating the changing periodicity of $\langle S^{x}_{0}S^{x}_{j}\rangle$.
Below $h_{s_{c}}$, the destructed quasi-LRO facilitates the staggered
magnetization and thus $\chi_{stag}$ increases. Above $h_{s_{c}}$, the
quasi-LRO is fully broken and $\chi_{stag}$ declines as in the Haldane
phase. Similar to the $S$=$1$ N\'{e}el phase, this transition is also
accompanied by the open of a gap, which, however, behaves in a different
manner above $h_{s_{c}}$
\begin{equation}
\Delta\sim(h_{s}-h_{s_{c}})^{\alpha}.  \label{fit2}
\end{equation}
The critical behaviors of $\Delta$ are fitted by Eq. (\ref{fit2}) in Fig. %
\ref{H-X}(d) with $\alpha$=$1.15$ and $1.1$ for $\beta$=$-3$ and $-5$,
respectively. These observations indicate that a QPT happens in $h_{s_{c}}$.
To characterize this QPT, the ground-state energy and its derivatives with
the field are studied. It is found that both $e$ and $\partial{e}/\partial{%
h_{s}}$ are nonsingular, but $\partial^{2}{e}/\partial{h_{s}}^{2}$ is
singular at $h_{s_{c}}$, indicating that this transition is of the
second-order, as shown in the inset of Fig. \ref{H-X}(d).

\begin{figure}[tbp]
\includegraphics[width = 1.0\linewidth,clip]{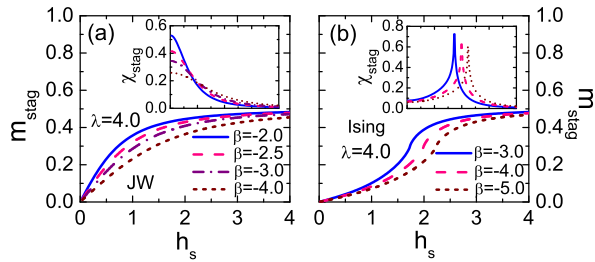}
\caption{(color online) Staggered magnetization and susceptibility obtained
by (a) the mean-field theory for the Hamiltonian Eq. (\protect\ref{HF}); and
(b) the exact solutions for the Hamiltonian Eq. (\protect\ref{Hamiltonian})
with only Ising interactions.}
\label{JW}
\end{figure}

As discussed above, the staggered transverse magnetic properties have
different behaviors in three phases. Although the QPT is observed in the $S$=%
$1$ N\'{e}el and XY-like phases, the physical quantities have different
behaviors in these phases. The QPT in the N\'{e}el phase is argued to be of
the KT type, while that in the XY-like phase is confirmed to be of the
second-order. For further discussions, the staggered magnetization is
studied in terms of the Jordan-Wigner (JW) transformation. As the field is
applied transversely, the transformation is introduced as
\begin{eqnarray}
S^{z}_{i}&=&\frac{1}{2}(c^{\dagger}_{i}e^{i\pi\sum_{j<i}c^{%
\dagger}_{j}c_{j}}+h.c.),  \nonumber \\
S^{y}_{i}&=&\frac{1}{2i}(e^{-i\pi\sum_{j<i}c^{\dagger}_{j}c_{j}}c_{i}-h.c.),
\nonumber \\
S^{x}_{i}&=&c^{\dagger}_{i}c_{i}-\frac{1}{2},
\end{eqnarray}
where $c^{\dagger}_{i}$ and $c_{i}$ are the creation and annihilation
operators of spinless fermions, which satisfy the anticommutation relation $%
\lbrace c_{i}, c^{\dagger}_{j}\rbrace $=$\delta_{ij}$. We denote the
fermions in odd and even sites as $a_{i}$ and $b_{i}$, respectively.
The density-density interaction terms are treated with the Hartree-Fock (HF)
approximation,
\begin{eqnarray}
n_{a,j}n_{b,j}&\approx&n_{a,j}\langle n_{b,j}\rangle+\langle n_{a,j}\rangle
n_{b,j}-(a^{\dagger}_{j}b_{j}\langle b^{\dagger}_{j}a_{j}\rangle+h.c.)
\nonumber \\
&+&(a^{\dagger}_{j}b^{\dagger}_{j}\langle b_{j}a_{j}\rangle+h.c.)-(\langle
n_{a,j}\rangle\langle n_{b,j}\rangle  \nonumber \\
&-&\langle b^{\dagger}_{j}a_{j}\rangle\langle a^{\dagger}_{j}b_{j}\rangle
+\langle b_{j}a_{j}\rangle\langle a^{\dagger}_{j}b^{\dagger}_{j}\rangle).
\end{eqnarray}
We denote $\langle a^{\dagger}_{i}a_{i}\rangle$=$n_{a}$, $\langle
b^{\dagger}_{i}b_{i}\rangle$=$n_{b}$, $\langle b_{i}a_{i}\rangle$=$p_{1}$, $%
\langle b^{\dagger}_{i}a_{i}\rangle$=$p_{2}$, $\langle
a^{\dagger}_{i+1}b_{i}\rangle$=$p_{3}$, and $\langle a_{i+1}b_{i}\rangle$=$%
p_{4}$. After making Fourier transform, the Hamiltonian is transformed into
\begin{eqnarray}
H_{HF}&=&\sum_{k}(\omega^{a}_{k}a^{\dagger}_{k}a_{k}+\omega^{b}_{k}b^{%
\dagger}_{k}b_{k})
+\sum_{k}(\omega^{1}_{k}a^{\dagger}_{k}b_{k}+\omega^{2}_{k}a^{%
\dagger}_{k}b^{\dagger}_{-k}  \nonumber \\
&+&h.c.)+const.,
\end{eqnarray}
where $\omega^{a}_{k}$=$(n_{b}-1/2)(\beta+1)-h_{s}$, $\omega^{b}_{k}$=$%
(n_{a}-1/2)(\beta+1)+h_{s}$, $\omega^{1}_{k}$=$[(\lambda+1)/4-p_{2}]e^{ik/2}$%
+$(1/2-p^{\ast}_{3})\beta e^{-ik/2}$, and $\omega^{2}_{k}$=$%
[(\lambda-1)/4+p_{1}]e^{ik/2}$-$p^{\ast}_{4}\beta e^{-ik/2}$. Then, we
introduce the Bogoliubov transformation
\begin{eqnarray}
a_{k}&=&u_{11}\alpha_{k}+u_{12}\beta_{k}+u_{13}\gamma_{k}+u_{14}\lambda_{k},
\nonumber \\
b_{k}&=&u_{21}\alpha_{k}+u_{22}\beta_{k}+u_{23}\gamma_{k}+u_{24}\lambda_{k},
\nonumber \\
a^{\dagger}_{-k}&=&u_{31}\alpha_{k}+u_{32}\beta_{k}+u_{33}\gamma_{k}+u_{34}%
\lambda_{k},  \nonumber \\
b^{\dagger}_{-k}&=&u_{41}\alpha_{k}+u_{42}\beta_{k}+u_{43}\gamma_{k}+u_{44}%
\lambda_{k}.
\end{eqnarray}
Therefore, the Hamiltonian is diagonalized as
\begin{equation}
H_{HF}=\sum_{k}(\omega^{\alpha}_{k}\alpha^{\dagger}_{k}\alpha_{k}+\omega^{%
\beta}_{k}\beta^{\dagger}_{k}\beta_{k}
+\omega^{\gamma}_{k}\gamma^{\dagger}_{k}\gamma_{k}
+\omega^{\lambda}_{k}\lambda^{\dagger}_{k}\lambda_{k}),  \label{HF}
\end{equation}
where $\omega_{k}^{i}$ ($i$=$\alpha,\beta,\gamma,\lambda$) are obtained by
the self-consistent numerical calculations. The staggered magnetic
properties are studied with this quardratic Hamiltonian. In the following,
we find that this mean-field theory is able to describe the behavior in the
Haldane phase, but fails to reproduce the QPT in other phases. Figure \ref%
{JW}(a) shows $m_{stag}$ and $\chi_{stag}$ from the Haldane to the N\'{e}el
phase. The analytic results qualitatively agree with the numerical ones in
the Haldane phase, but cannot find the inflexion in the N\'{e}el phase.
However, some exactly soluble case that has a competition between the field
and magnetic couplings also shows QPT and may be compared with our
observations. Figure \ref{JW}(b) shows the exact solutions for the Ising
case of Eq. (\ref{Hamiltonian}), which exhibits a QPT with increasing $h_{s}$%
. The transition behaviors are analogous to those in the XY-like phase. The
singular $\partial^{2}{e}/\partial{h_{s}}^{2}$ in $h_{s_{c}}$ indicates that
this QPT also belongs to the second-order, which are not presented here. The
gap above $h_{s_{c}}$ behaves as $\Delta$$\sim$$(h_{s}-h_{s_{c}})$, which is
Eq. (\ref{fit2}) with $\alpha$=$1$. It may be expected that the QPT in the
Ising case is similar to that in the $S$=$1$ N\'{e}el phase, but it is of
the second-order, like the transition in the XY-like phase. This phenomenon
shows the important role of the xy-component quantum fluctuations in the
QPT, which \textit{smooth} the transition in the N\'{e}el phase.

In summary, we have studied the magnetic properties of the spin-$1/2$ AF-FM
Heisenberg chain with AF anisotropy in a transverse staggered magnetic field
by means of the DMRG method. The physical quantities are explored in the
Haldane, $S$=$1$ N\'{e}el, and XY-like gapless phases. In the Haldane phase,
$m_{stag}$ and $\chi_{stag}$ behave as those of the $S$=$1$ Haldane chain,
and do not have transition behaviors. In the N\'{e}el phase and the XY-like
gapless phase, due to the competitions between the field and different
magnetic couplings, most quantities have a transition at a field $h_{s_{c}}$%
, indicating a QPT induced by the field happens in the system. $m_{stag}$
has an inflexion in $h_{s_{c}}$, which corresponds to a maximum in $%
\chi_{stag}$. The transition is also accompanied by the open of a gap. In
the N\'{e}el phase, $\partial^{2}{e}/\partial{h_{s}}^{2}$ is nonsingular,
but it still describes the transition with a minimum at the critical field.
The QPT is argued to be of the KT type. The critical behavior of $\Delta$ is
well fitted by that in the KT transition near $h_{s_{c}}$. In the XY-like
gapless phase, $\partial^{2}{e}/\partial{h_{s}}^{2}$ is singular in $%
h_{s_{c}}$, indicating that the transition is of the second order. The gap
near $h_{s_{c}}$ behaves as $\Delta$$\sim$$(h_{s}-h_{s_{c}})^{\alpha}$. Due
to the distinct magnetic orders, the transitions behave differently in the
two phases. Using the Jordan-Wigner transformation, the magnetic properties
are also investigated analytically. In the present HF approximation, the
features in the Haldane phase are reproduced, but it fails to describe the
transitions in other two phases. The Ising case of the Hamiltonian (\ref%
{Hamiltonian}) is exactly resolved, which has a second-order QPT that
behaves like the one in the XY-like phase. The differences between the Ising
case and the N\'{e}el phase show the important role of the quantum
fluctuations.

\acknowledgments

We are grateful to Wei Li and Yang Zhao for useful discussions. This work is
supported in part by the National Science Fund for Distinguished Young
Scholars of China (Grant No. 10625419), the National Science Foundation of
China (Grants No. 90403036 and No. 20490210), the MOST of China (Grant No.
2006CB601102), and the Chinese Academy of Sciences.

\end{document}